\documentclass[
reprint,
superscriptaddress,
amsmath,amssymb,
aps,
twocolumn,
floatfix,
longbibliography
]{revtex4-2}

\usepackage{graphicx}
\usepackage{dcolumn}
\usepackage{bm}
\usepackage{graphicx}
\usepackage{currfile}
\usepackage{multirow}
\usepackage{chemformula}
\usepackage{bm,soul}
\usepackage{color}
\usepackage{gensymb}
\usepackage{amsmath}
\usepackage{amssymb}
\usepackage{dcolumn}
\usepackage{comment}

\newcommand{\Ang}{\AA$^{-1}$}

\newcommand{\XQE}{\ensuremath{\chi''(Q,E)}}

\definecolor{grey}{gray}{.5}

\definecolor{malacol}{rgb}{0.12,0.69,0.44}

\newcommand{\uB}{\ensuremath{\mu_\textrm{B}}}

\newcommand{\barlo}{Cu$_4$(OH)$_{6}$FBr}

\newcommand{\clarid}{Cu$_4$(OD)$_{6}$FCl}

\newcommand{\znclarid}{ZnCu$_3$(OD)$_{6}$FCl}

\newcommand{\xclarid}{Zn$_x$Cu$_{4-x}$(OD)$_{6}$FCl}

\begin{document}

\title{Magnetically ordered and kagome quantum spin liquid states in the Zn-doped claringbullite series}

\author{M.~Georgopoulou} 
\affiliation{Institut Laue-Langevin, CS 20156, 38042 Grenoble Cedex 9, France}
\affiliation{Department of Chemistry, University College London, 20 Gordon Street, London, WC1H 0AJ, United Kingdom}

\author{B.~F\aa k}
\email{fak@ill.fr}
\affiliation{Institut Laue-Langevin, CS 20156, 38042 Grenoble Cedex 9, France}

\author{D. Boldrin} 
\affiliation{SUPA, School of Physics and Astronomy, University of Glasgow, Glasgow, G12 8QQ, United Kingdom}

\author{J. R. Stewart} 
\affiliation{ISIS Neutron and Muon Facility, Rutherford Appleton Laboratory, Science and Technology Facilities Council, Didcot OX11 0QX, United Kingdom}

\author{C. Ritter} 
\affiliation{Institut Laue-Langevin, CS 20156, 38042 Grenoble Cedex 9, France}

\author{E. Suard} 
\affiliation{Institut Laue-Langevin, CS 20156, 38042 Grenoble Cedex 9, France}

\author{J.~Ollivier} 
\affiliation{Institut Laue-Langevin, CS 20156, 38042 Grenoble Cedex 9, France}

\author{A.~S.~Wills} 
\affiliation{Department of Chemistry, University College London, 20 Gordon Street, London, WC1H 0AJ, United Kingdom}

\date{\today \quad} 

\begin{abstract}
Neutron scattering measurements have been performed on deuterated powder samples of claringbullite and Zn-doped claringbullite (\xclarid). 
At low temperatures, claringbullite \clarid\ forms a distorted pyrochlore lattice with long-range magnetic order and spin-wave-like magnetic excitations. 
Partial Zn doping leads to the nominal \znclarid\ compound, a geometrically frustrated spin-1/2 kagome antiferromagnet that shows no transition to magnetic order down to 1.5~K. 
The magnetic excitations form a gapless continuum, a signature of fractional excitations in a quantum spin liquid.
\end{abstract}

\maketitle

\section{Introduction}
\label{SecIntro}
Interacting spins forming a two-dimensional lattice of corner sharing triangles, the kagome lattice, are expected to form unconventional magnetic ground states due to geometric frustration of the exchange interactions. 
These ground states are predicted to include zero-temperature long-range ordered states with non-collinear and even non-coplanar multi-{\bf k} magnetic orders for classical spins \cite{Messio2011}. 
For quantum spins, in particular spin-1/2, a large variety of more exotic quantum spin liquid (QSL) ground states have been predicted \cite{Balents2010,Bieri2015,Bieri2016}. 
A crucial question for the classification of a spin liquid is whether the excitation spectrum has a gap or not \cite{Broholm2020QuantumLiquids}. 

\begin{figure}[!b]
\centering
\includegraphics[width=0.6\columnwidth]{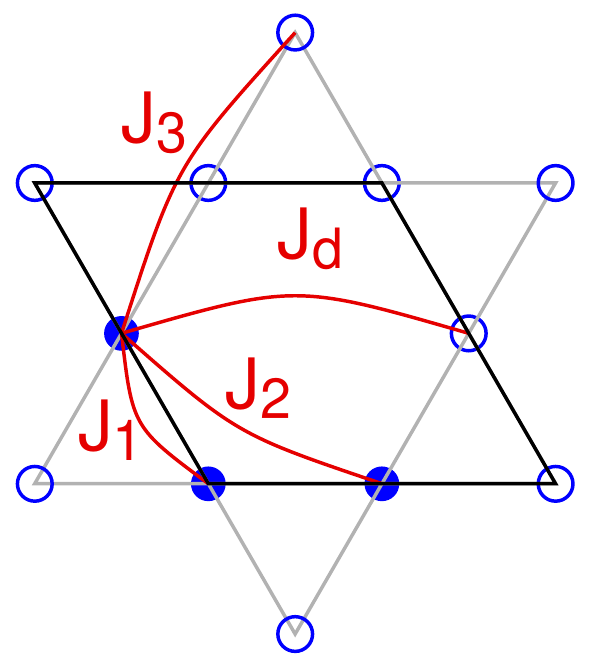}
\caption{Kagome lattice showing the first four exchange interactions. 
Black lines show the unit cell with three Cu$^{2+}$ $S=1/2$ ions as filled circles. 
}
\label{FigKagome}
\end{figure}

Despite much effort, there are relatively few good experimental model systems to study spin-1/2 kagome QSLs. 
One example is the mineral herbertsmithite, ZnCu$_3$(OH)$_{6}$Cl$_{2}$, 
where the magnetic response resembles a collection of uncorrelated nearest-neighbor (NN) singlets \cite{Han2012}. 
The existence of a gap for this state is currently under discussion \cite{Fu2015,Khuntia2020}, chiefly because of Cu/Zn mixing on the Cu site. 
Such site-disorder has also hampered the study of the $\alpha$ polymorph of herbertsmithite, the mineral kapellasite \cite{Fak2012,Kermarrec2014}, 
where a competition between NN ferromagnetic $J_1<0$ and an antiferromagnetic 3rd-neighbor interaction $J_d$  \cite{Fak2012,Bernu2013} (see Fig.~\ref{FigKagome} for a definition of the exchanges) leads to a gapless chiral quantum spin liquid within the cuboc2 part of the classical phase diagram \cite{Messio2011,Bieri2015}. 

In this work, we study another candidate kagome QSL, \znclarid\ (hereafter referred to as ZnCu3), 
obtained by partial Zn-doping of the mineral claringbullite \clarid\ (hereafter referred to as Cu4). 
In the latter, three of the four Cu$^{2+}$ ions in the unit cell form perfect spin-1/2 kagome layers at room temperature. 
The fourth Cu$^{2+}$ ion sits alternatively above and below the kagome triangles, leading to a slightly distorted pyrochlore lattice. 
The resulting distorted trigonal prism coordination of the interplanar Cu$^{2+}$ ions contrasts with the distorted octahedral coordination of herbertsmithite.
This coordination geometry is expected to drastically reduce Cu/Zn site disorder in ZnCu3,  
since the closed-shell Zn$^{2+}$ ion prefers the interlayer site compared to the more distorted octahedral sites of the kagome plane, as predicted in the isostructural mineral barlowite \barlo\ \cite{Liu2015}. 

Claringbullite (Cu4), first reported with a slightly different chemical formula \cite{Fejer1977}, 
crystallizes at room temperature in the hexagonal space group $P6_3/mmc$ (No.~194) \cite{Burns1995}. 
The interlayer Cu, which is disordered over three equivalent positions in this structure, orders slightly below room temperature, leading to an orthorhombic distortion and a doubling of the crystallographic unit cell \cite{Henderson2019,Tustain2021}. 
The magnetic interactions in claringbullite are predominantly antiferromagnetic, as seen from the Weiss temperature $\theta_\mathrm{W}=-135$~K, and magnetic order is established below $T_\mathrm{N}=17$~K \cite{Tustain2021,Yue2018,Feng2018b}. 
Zinc-doped claringbullite (ZnCu3), on the other hand, remains in the $P6_3/mmc$ space group down to the lowest temperatures, as will be shown below. 
It does not order magnetically down to at least $T=0.8$~K according to specific heat measurements  despite a largely antiferromagnetic $\theta_\mathrm{W}$ of $-223$~K \cite{Feng2018b}, and is therefore a promising kagome QSL candidate. 
Here, we present inelastic neutron scattering measurements on highly deuterated \clarid\ and \znclarid\ powder samples, and their crystallographic and magnetic characterization.

\section{Synthesis and Characterization}
\label{SecExp}
\subsection{Synthesis}
\label{SecSynthesis}
Highly deuterated Cu4 and ZnCu3 polycrystalline samples of weight 5--7~g were synthesized using a hydrothermal method adapted from \cite{Feng2018b}, as detailed in the Supplemental Material \cite{Supplemental}.

\subsection{Crystallographic studies}
\label{SecStructure}
Laboratory X-ray powder diffraction ($\lambda=1.5406$~\AA) showed both the Cu4 and ZnCu3 samples to be single phase and to crystallize in the $P6_3/mmc$ space group at room temperature. 
This was confirmed by high-resolution powder neutron diffraction using D2B at the ILL, where measurements were performed with $\lambda=1.5952$~\AA\ at $T=1.5$ and 295~K \cite{D2B_cbull}. 
Rietveld refinements showed a high deuteration level of $\sim$96\%. 

At $T=1.5$~K, the powder neutron diffraction data from Cu4 show additional peaks that were successfully indexed in the $Pnma$ space group ($a=11.5359(9)$, $b=9.1510(7)$, and $c=6.6848(5)$~\AA), in agreement with previous single crystal studies at $T=100$~K \cite{Henderson2019}. 
The driving force for this distortion is the ordering of the interlayer Cu onto one of three local equivalent sites. At low temperatures, the Cu4 $Pnma$ structure is a distorted pyrochlore with three Cu sites at the $4a$, $8d$ and $4c$ Wyckoff positions. The kagome to interlayer Cu distances range from 2.80 to 3.17~\AA, slightly smaller than the kagome Cu-Cu distances that range from 3.32 to 3.35~\AA. 

ZnCu3 does not show additional peaks at $T=1.5$~K, indicating that it remains in the $P6_3/mmc$ space group ($a=6.65918(6)$ and $c=9.17288(9)$~\AA).
The refinement of the low temperature neutron diffraction data shows that the kagome site is within the error of the refinement fully occupied by Cu. The interlayer position is occupied by $\sim$74\% Zn on the high-symmetry $2d$ site while the Jahn-Teller active Cu ion occupies $\sim$8.5\% of the $6h$ site (see Supplemental Material \cite{Supplemental}).

\subsection{Magnetometry}
\label{SecSquid}

Magnetic susceptibility measurements in a magnetic field of 1000~G showed a magnetic transition in Cu4 at $T_\mathrm{N}\approx17$~K with a weak ferromagnetic component (possibly due to Dzyaloshinskii-Moriya interactions) and a Weiss temperature of $\theta_\mathrm{W}=-136(3)$~K (see Supplemental Material \cite{Supplemental}). In ZnCu3, no transition to magnetic order was seen down to 2~K and the Weiss temperature was $\theta_\mathrm{W}=-206(1)$~K (Supplemental Fig.~S5 \cite{Supplemental}), in agreement with \cite{Feng2018b}. 

Magnetization measurements as a function of magnetic field show, 
contrary to the literature \cite{Feng2018b}, a small hysteresis in ZnCu3 of 0.06~T with a spontaneous moment of $2\times10^{-4}\,\uB$/Cu at $T=2$~K, which corresponds to 0.01\% of the full ordered moment (see Supplemental Fig.~S5 \cite{Supplemental}). This could be intrinsic to our ZnCu3 sample, however, we note that the hysteresis loop opens below 6~K, which corresponds to the ordering temperature of clinoatacamite \cite{Wills2008}. Although no impurity phase was observed in the D2B neutron diffraction data, previous work has reported impurities of polymorphs of clinoatacamite \cite{Tustain2021} or herbertsmithite \cite{Feng2018FromFCl} in both Cu4 and ZnCu3. 
The observed spontaneous moment at 2~K could be attributed to a 0.4\% clinoatacamite impurity \cite{zheng2005}, which would not be visible in our powder diffraction data.

\begin{figure}
\centering
\includegraphics[width=0.99\columnwidth, trim=4 4 4 4,clip]{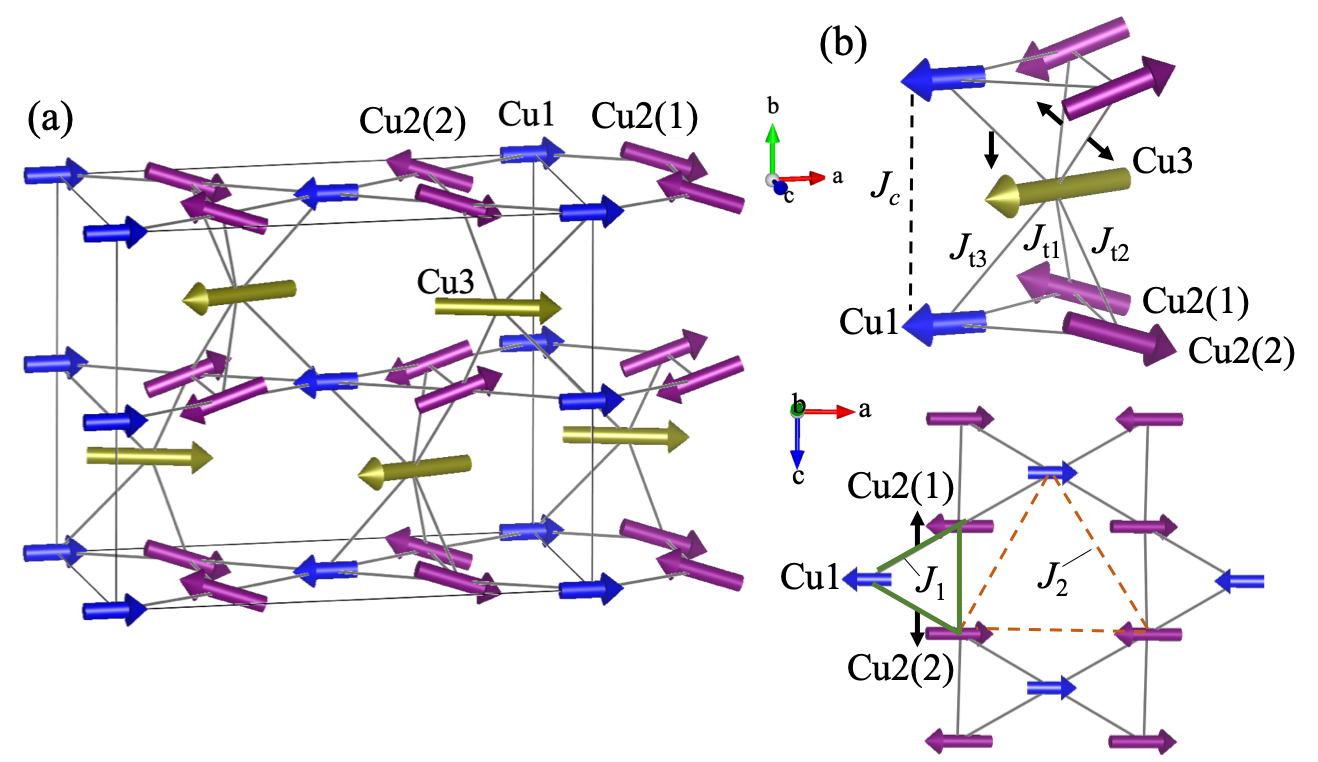}
\caption{(a) Magnetic structure of \clarid\ at $T=1.7$~K. (b)~Relevant exchange interactions in the orthorhombic structure. The black arrows indicate the Dzyaloshinskii-Moriya vectors along the [010] and [001] directions (see Table \ref{TableExchanges}). 
}
\label{FigMagStruc}
\end{figure}

\section{Magnetic structure}
\label{SecMag}
The magnetic structure of Cu4 was determined at $T=1.7$~K using the high-flux neutron diffractometer D20 at the ILL with $\lambda=2.410$~\AA\ \cite{D20_cbull}.
Subtraction of data taken at $T=25$~K, just above the magnetic phase transition at $T_\mathrm{N}=17$~K, show more than a dozen magnetic peaks that correspond to an antiferromagnetic order with propagation vector ${\bf k}=0$ in the orthorhombic space group (see Supplemental Material \cite{Supplemental}).
Refinements of the different irreducible representations (IRs) obtained by representation analysis \cite{Wills2000ASARAh} gave a good fit only for the IR $\Gamma_7$, in Kovalev's notation, 
which corresponds to the magnetic space group $Pn^\prime m^\prime a$, previously found for both barlowite \cite{Tustain2021,Tustain2018} and claringbullite \cite{Tustain2021}. 
The ordered magnetic moments are found to be reduced from the fully ordered 1~\uB\ expected for $S=1/2$, which we relate to quantum fluctuations, with values of  0.26, 0.37, and 0.52~\uB\ for the Cu1 site at (0,~0,~0), Cu2 at (0.247,~0.496,~0.249), and Cu3 at (0.191,~1/4,~0.051), respectively. Details of the basis vectors for the IR $\Gamma_7$ can be found in the Supplemental Material \cite{Supplemental}. 

The magnetic moments are mostly aligned along the orthorhombic $a$ axis, 
with the Cu2 moment rotated (antiferromagnetically) by 18$^\circ$ towards $b$ 
and the Cu3 moment canted ferromagnetically by 17$^\circ$ towards $c$, 
see Fig.~\ref{FigMagStruc}(a). 
Of the three spins on each kagome triangle, two are almost parallel to the $a$-axis and so to the interlayer spin and one is antiparallel. The former are closer in distance to the interlayer spin, and one may speculate that the  ``tripod'' interactions $J_{tn}$ ($n=1,2,3$) between the interlayer and the kagome spins are ferromagnetic,  
in agreement with density functional theory calculations for barlowite \cite{Jeschke2015}. 
This scenario is also supported by the smaller negative value of the Weiss temperature of Cu4 compared to ZnCu3, which suggests additional ferromagnetic interactions in the former.
The canting of the spins away from the $a$ axis indicates that antisymmetric Dzyaloshinskii-Moriya (DM) interactions may play a role. 
Measurements at $T=15$~K, just slightly below $T_\mathrm{N}$, show the same magnetic structure as at $T=1.7$~K, with further reduced moment sizes due to thermal fluctuations. 

The magnetic structure found for Cu4 in this work is similar to that reported for a different polycrystalline sample \cite{Tustain2021}, with the main difference being 20-30\% smaller moment sizes in our samples.
The magnetic structure also resembles that of barlowite, which has slightly larger canting angles: $\sim$28$\degree$ for Cu2 and $\sim$22$\degree$ for Cu3 \cite{Tustain2018}.

\section{Inelastic neutron scattering}
\label{SecINS}
\subsection{Measurements}
\label{SecSpectro}
Inelastic neutron scattering (INS) measurements on ZnCu3 were performed on the cold-neutron time-of-flight (TOF) spectrometer LET (ISIS) using neutrons with incoming energies of $E_i=2.8,~6.0$ and 20~meV at temperatures of 1.7 and 50~K \cite{LET_cbull}, while
Cu4 was measured on the cold and thermal TOF spectrometers IN5 and Panther (ILL) 
using incoming neutron energies between 3.5 and 35~meV at temperatures between 1.7 and 100~K \cite{Panther_cbull, IN5_cbull}. 
Standard data reduction was carried out using MANTID \cite{Mantid}. For comparison between the two samples, the scattering was put on an absolute scale by normalizing to the intensities of nuclear Bragg peaks.

\begin{figure}
\includegraphics[width=1\columnwidth, trim=4 4 4 4,clip]{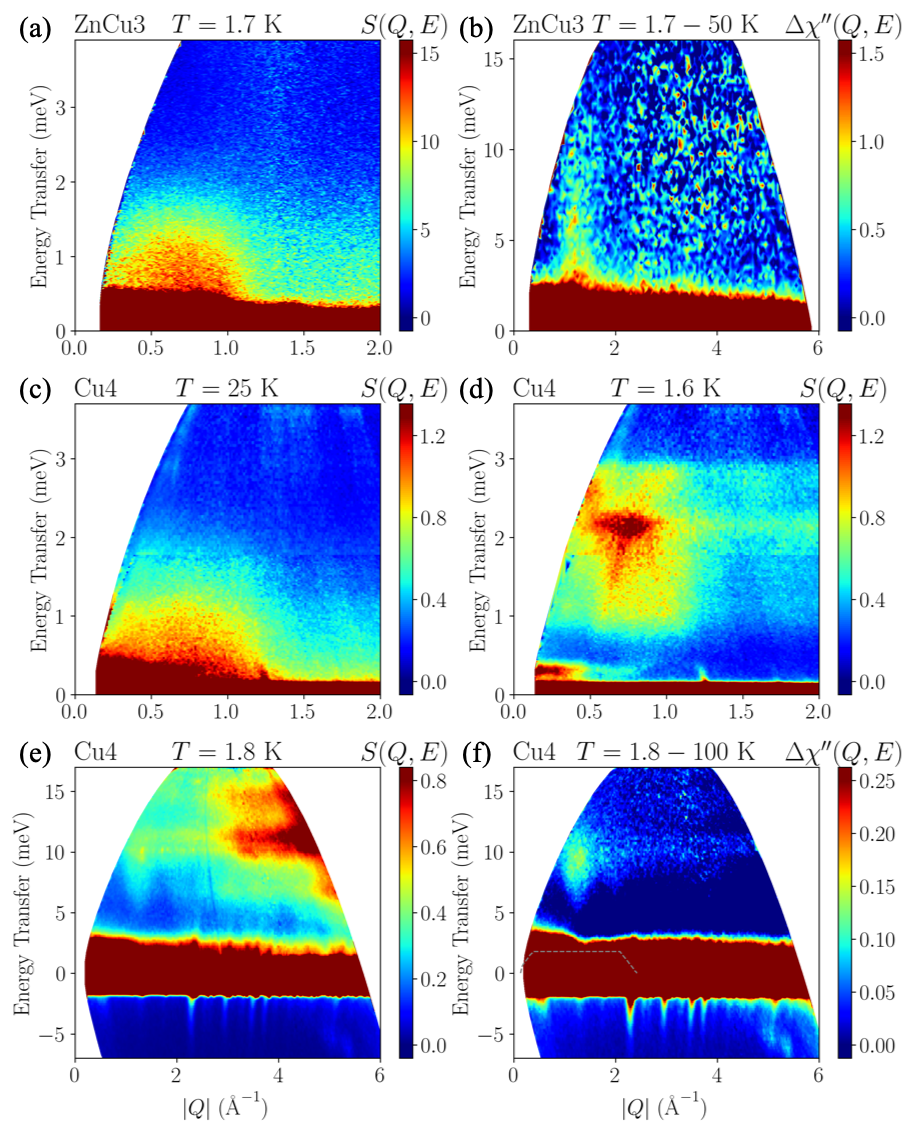}
\caption{
(a--b)~INS spectra from \znclarid\ at $T=1.7$~K measured on LET:
(a)~$S(Q,E)$ for $E_i=6.0$~meV;
(b)~Difference in magnetic dynamic susceptibility $\chi^{\prime\prime}(Q,E)$ between $T=1.7$ and 50~K for $E_i=20$~meV. 
(c--d)~$S(Q,E)$ of \clarid\ at two different temperatures measured on IN5 by combining $E_i = 3.55$ and 5.1~meV:
(c)~Cooperative paramagnetic state at $T=25$~K;
(d)~Magnetically ordered state at $T=1.6$~K. 
(e--f)~INS spectra from \clarid\ measured on Panther with $E_i=19$~meV: 
(e)~$S(Q,E)$ at $T=1.8$~K; 
(f)~Difference in magnetic dynamic susceptibility $\chi^{\prime\prime}(Q,E)$ between $T=1.8$ and 100~K showing both the low-energy response in the tail of the elastic peak and the high-energy response near 10~meV.
All data in this figure are in arbitrary units.
}
\label{FigMaps}
\end{figure}

\subsection{Zn-doped claringbullite \znclarid}
\label{SecZnCu3}
No magnetic Bragg peaks were observed in data collected for ZnCu3 at $T=1.7$~K, which confirms the absence of magnetic order in this compound. 
The scattering function $S(Q,E)$ at $T=1.7$~K shows a low-energy spin-liquid-like magnetic response centered at about $Q=0.7$~\Ang\ with an energy-independent full width at half maximum in $Q$ of about 0.7~\Ang, see Fig.~\ref{FigMaps}(a). 
The excitations are peaked at about 1~meV with a non-Lorentzian energy profile
and a band width of about 3~meV. 
The excitations have no discernible energy gap, at least down to 0.27~meV. 

In addition to the clear low-energy magnetic response, there is a second contribution at higher energies that also appears to be magnetic. 
It is centered at $Q\approx1.2$~\Ang\ [see Fig.~\ref{FigMaps}(b)] and has a typical width of about 0.7~\Ang, 
which corresponds to 
a correlation length of 8.4~\AA\ (approximately 1.2 kagome hexagons). 
This excitation has a band width of about 10~meV and no discernible gap. 

These two magnetic responses have different energy ranges  but can both be indexed by a $\textbf{k}=\textbf{0}$ characteristic wave vector, corresponding to (0,0,1) and (1,0,1) reflections, respectively. 
At $T=50$~K, both magnetic responses persist but are weaker in intensity.

\subsection{Claringbullite \clarid}
\label{SecCu4}
In Cu4 above the ordering temperature, a low-energy magnetic response is seen at $Q=0.7$~\Ang\ [see Fig.~\ref{FigMaps}(c)], very similar to that observed in ZnCu3. 
In the magnetically ordered phase, this scattering develops features strongly resembling dispersive spin waves, see Fig.~\ref{FigMaps}(d), with an energy gap of 0.45~meV and a band width of 3.35~meV. Additionally, there is a high-energy magnetic response centered at $Q\approx1.2$~\Ang, with a 5~meV energy gap and scattering extending up to 13~meV [see Figs.~\ref{FigMaps}(e) and \ref{FigMaps}(f)].

\section{Analysis}
\label{SecAna}

\subsection{Spin waves in \clarid}
\label{SubSecSpinWaves}
An attempt was made to describe the observed spin waves of Cu4 using semi-classical linear spin wave theory with the spin Hamiltonian
\begin{align}
    \mathcal{H} = \sum_{i,j} J_{ij} \mathbf{S}_i \cdot \mathbf{S}_j + \mathbf{D}_{ij} \sum_{i,j} \mathbf{S}_i \times \mathbf{S}_j \,.
\label{EqnCbullHam}
\end{align}
It turned out to be quite difficult to stabilize the observed magnetic structure, 
even when including isotropic exchange interactions $J_{ij}$ up to six further neighbors as well as
antisymmetric anisotropic DM interaction terms $\mathbf{D}_{ij}$. 
The exchange interactions predicted from combined density functional theory calculations and magnetic susceptibility measurements for barlowite \cite{Jeschke2015}, which has similar crystallographic and magnetic structures to claringbullite, also did not stabilize the observed magnetic structure in claringbullite. 

The complicated magnetic structure in Cu4 suggests that several exchange interactions are involved. To explore the sizable phase space of Eq.~(\ref{EqnCbullHam}), the program Serendipity \cite{Serendipity} was developed to identify cluster areas of phase space where the experimental magnetic structure could be stabilized by a given set of exchange parameters. The magnetic excitation spectra from the resulting sets of exchange integrals were calculated using SpinW \cite{Toth2015LinearStructures}, allowing for a relaxation of the observed canting angles in the magnetic structure to improve stability, and compared to the experimental data. 

The final model includes the exchange interactions $J_{ij}$ shown in Fig.~\ref{FigMagStruc}(b) and specified in Table~\ref{TableExchanges}, namely: 
a strong antiferromagnetic nearest-neighbor interaction $J_1$ in the kagome plane, 
three different ferromagnetic ``tripod'' interactions $J_{t1}$, $J_{t2}$, and $J_{t3}$ between the kagome spins and the Cu3 spin that caps the triangles, 
an antiferromagnetic interaction $J_c$ between the kagome planes, 
and a very weak antiferromagnetic next-nearest interaction $J_2$ in the kagome plane. 
This set of exchange interactions give a Weiss temperature of  $\theta_\mathrm{W}=-95$~K,  
in quite close agreement with the experimental value of $\theta_\mathrm{W}=-136$~K, 
and the signs of the interactions are compatible with the expectations from the superexchange bond angles. 
DM interactions (with both in- and out-of-plane components) were added to the $J_1$, $J_{t1}$ and $J_{t2}$ exchange paths to stabilize the spin canting of the magnetic structure. 
The calculated spin wave spectra, shown in the Supplemental Material \cite{Supplemental}, indicate that the model correctly reproduces the bandwidth of the observed spin waves, including the zero-energy gap and the gap between the upper and lower branches. 
The intensity ratio between these two branches is also reproduced, as well as the overall $Q$ dependence, but the details of the excitation spectra differ substantially.  
Further progress would require measurements on large single crystals, currently unavailable. 

\begin{table}
\caption{Exchange interactions used for the spin-wave calculations. 
The labels refer to Fig.~\ref{FigMagStruc}(b) and positive values denote antiferromagnetic couplings. Dzyaloshinskii-Moriya interactions along the [010] and [001] directions are given.}

\begin{ruledtabular}
\begin{tabular}{cccc}
Label & $J_i$ (meV)  & $|{D}_{[010]}|/J_i$ & $|{ D}_{[001]}|/J_i$\\
\hline
$J_1$ & 14.0 & 0 & 0.03 \\
$J_{t1}$ & -6.5 & 0 & 0.03 \\
$J_{t2}$ & -3.5 & 0 & 0.03 \\
$J_{t3}$ & -5.9 & 0.05 & 0 \\
$J_{c}$ & 4.0 & 0 & 0 \\
$J_{2}$ & 0.2 & 0 & 0 \\
\end{tabular}
\end{ruledtabular}

\label{TableExchanges}
\end{table}

\subsection{Zeroth moment analysis of \znclarid}
\begin{figure}[!t]
\centering
\includegraphics[width=0.99\columnwidth, trim=13 4 4 4,clip]{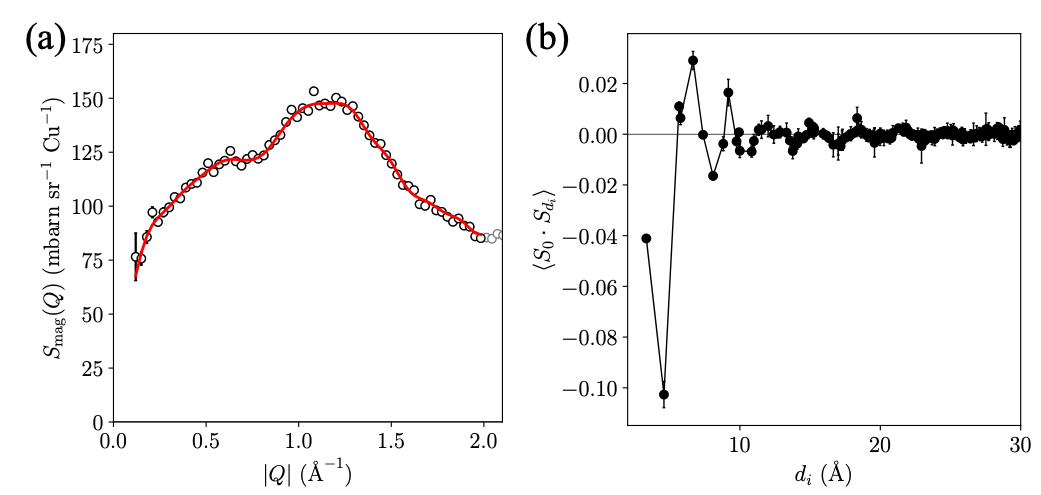}
\caption{
(a)~$S_{\rm mag}(Q)$ of \znclarid\ at $T=1.7$~K from measurements on LET with $E_i=2.8,~6.0$ and 20~meV (black symbols) with RMC fit (red line).
(b)~Radial spin correlations from RMC fits as a function of Cu-Cu distance $d$.
}
\label{FigSpinvert}
\end{figure}

Qualitative information on spin correlations can be obtained using zeroth moment analysis of the neutron scattering intensity $S(Q,E)$. 
This method was applied to ZnCu3 data taken at $T=1.7$~K with incoming energies of 2.8, 6.0 and 20~meV. 
The zeroth moment of the magnetic scattering, $S_{\rm mag}(Q)$, was obtained by integrating the measured (and normalized) $S(Q,E)$ over the energy range $0.1<E<12$~meV. 
To reduce the influence of coherent phonon scattering, which is important for wave vectors $Q>2$~\Ang, the wave-vector range was limited to below that value, and the data were fed into the SPINVERT program \cite{Paddison2013Spinvert:Data}, which uses reverse Monte-Carlo (RMC) to extract radial spin correlations $\langle {\bf S}_0 \cdot  {\bf S}_d \rangle$.
The data of $S_{\rm mag}(Q)$ and the RMC fit are shown in Fig.~\ref{FigSpinvert}(a).
The refined effective magnetic moment, $\mu_\mathrm{eff} = 1.705(1)$~\uB, is in good agreement with the spin-only value of 1.73~\uB.
The radial spin correlations are shown as a function of Cu--Cu distance $d$ in Fig.~\ref{FigSpinvert}(b).
The strongest correlation is negative and corresponds to the interkagome distance, $\sim$4.56~\AA, indicating predominantly antiparallel spin alignment between kagome layers. The nearest-neighbor correlation in the kagome plane ($d\approx3.33$~\AA) is also antiferromagnetic, as expected from the large negative Weiss temperature.

\subsection{Comparison between Cu4 and ZnCu3}
\label{SecComp}
\begin{figure}[!t]
\centering
\includegraphics[width=0.98\columnwidth, trim=4 0 2 4,clip]{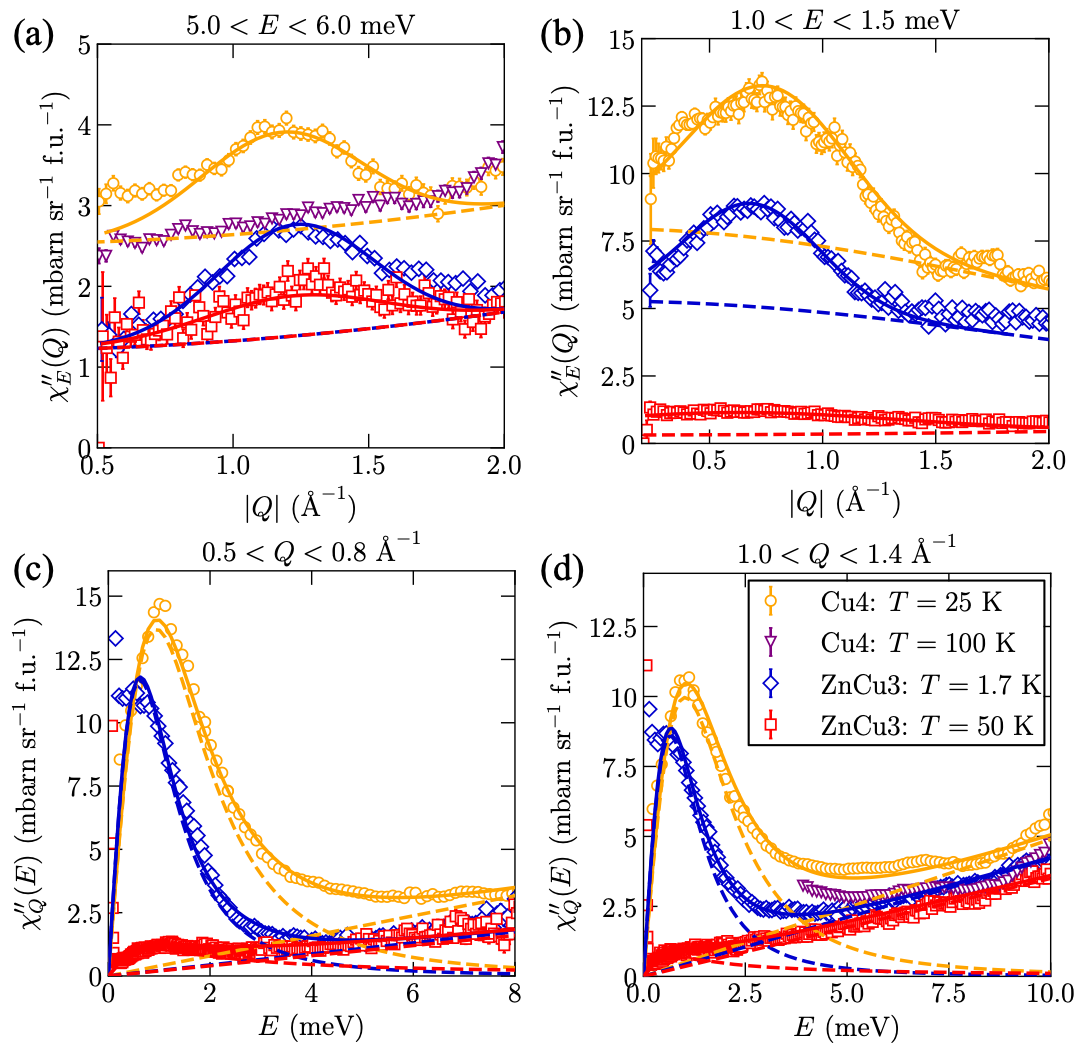}

\caption{Comparison of scans in $Q$ [(a)--(b)] and energy [(c)--(d)] of the dynamic magnetic susceptibility in different energy and $Q$ ranges, respectively, of \znclarid\ at $T=1.7$~K (blue) and 50~K (red) and of \clarid\ at $T=25$~K (orange) and $T=100$~K (purple) showing the similarities between the two systems. The solid lines are fits as described in the text. The data are normalized to the number of formula units. 
}
\label{FigScans}
\end{figure}

In the cooperative paramagnetic regime, 
the magnetic scattering shows very strong resemblances between ZnCu3 at low temperature and Cu4 above $T_\mathrm{N}$, as shown in Fig.~\ref{FigMaps}(a) and (c).
This is further illustrated in Fig.~\ref{FigScans} by cuts of $\chi^{\prime\prime}(Q,E)$ in both constant energy and $Q$. 
Scans at constant $Q$ were fitted with three contributions:
a Gaussian describing magnetic correlations, 
a $Q$-independent term describing uncorrelated magnetic scattering, 
and an incoherent phonon term given by $a+bQ^2\exp(-u^2Q^2/2)$, where $a$ is due to multiple scattering. 

At energies of about 5--6~meV, Fig.~\ref{FigScans}(a),
the correlated magnetic scattering is centered at $Q\approx1.2$~\Ang\ in both samples, with similar widths in $Q$, indicating similar correlations in both materials. 
The scattering underneath these correlations is dominated by incoherent phonons. 
This phonon scattering is higher in the Cu4 sample, possibly due to a lower deuteration level, corresponding to 90\% D instead of the 96\% obtained from the refinement of the diffraction data.
The peak areas of Cu4 at $T=25$~K and ZnCu3 at $T=1.7$~K are within 10\% of each other, suggesting that the number of Cu atoms contributing to the higher energy response is the same. 
This gives strong evidence for the magnetic scattering near $Q=1.2$~\Ang\ being due to correlations from the kagome layers. 
These correlations persist up to at least $T=50$~K and possibly vanish at $T=100$~K.

At lower energies of $\sim$1--1.5~meV, Fig.~\ref{FigScans}(b), the correlated magnetic scattering is centered at $Q\approx 0.7$~\Ang\ in both samples, with similar widths in $Q$, again indicating similar correlations in both materials. 
The broad scattering beneath the correlations is predominantly attributed to uncorrelated magnetic scattering and the remainder arises from incoherent phonons.
Both the correlated and uncorrelated magnetic contributions are about 1.5 times stronger in Cu4 than in ZnCu3. 
This suggests that it is the same atoms that contribute to the correlated and uncorrelated magnetic scattering, and that the scattering comes from all Cu atoms in each sample.

The analysis of neutron diffraction data collected from ZnCu3 indicates $\sim$8.5\% occupancy of the interlayer $6h$ sites by Cu$^{2+}$. 
This occupation by a moment bearing ion would introduce some local `tripod' exchange.
Comparison of the $Q$ dependences of the magnetic scattering of Cu4 ($T=25$~K) and ZnCu3 ($T=1.7$~K) shown in Fig.~\ref{FigScans}, suggests that in this regime the `tripod' exchange  in these materials does not notably change the magnetic excitations. 

The energy dependence of the excitations was compared by integrating \XQE\ over wave vector ranges $0.5<Q<0.8$~\Ang\ and $1.0<Q<1.4$~\Ang, shown in Figs.~\ref{FigScans}(c) and \ref{FigScans}(d), respectively.
The low-energy peak is not well described by a quasielastic Lorenzian corresponding to a single relaxation rate, but can be fitted with a squared quasielastic Lorentzian function, $\chi''(E) = 2E \chi^\prime \Gamma^3/[(E^2+\Gamma^2)^2]$, describing an overdamped excitation where $\Gamma$ is related to the width in energy of the scattering. 
An additional contribution, linear in energy, $\chi''(E)~=~eE$, had to be added to the fits to describe the high-energy behavior. 
This could arise from both magnetic and phonon scattering, but its very weak temperature dependence indicates that it mainly arises from the density of states of acoustic phonons with energies below the Debye temperature.

At low $Q$, Fig.~\ref{FigScans}(c), Cu4 at $T=25$~K has a maximum at $E=\Gamma/\sqrt{3}=0.940(3)$~meV, whereas ZnCu3 peaks at $0.624(2)$~meV at $T=1.7$~K. 
At all $Q$ values, the low-energy magnetic response of Cu4 has a larger $\Gamma$ than ZnCu3, with values of 1.628(5)~meV and 1.081(3)~meV, respectively, from Fig.~\ref{FigScans}(c).

\section{Discussion}
The model-free zeroth-moment analysis of ZnCu3 indicates that the strongest correlations are antiferromagnetic and between the kagome layers. 
Therefore it is likely that these are the main contribution to the low-energy magnetic scattering in ZnCu3, as also supported by the observation that the scattering in Fig.~\ref{FigScans}(b) is centered at $Q=0.7$~\Ang, corresponding to the (0,0,1) reflection of the $\textbf{k} = \textbf{0}$ wave vector in reciprocal space.
The low-energy excitation of Cu4 and ZnCu3 have very similar profiles with Cu4 showing additional intensity compared to ZnCu3 in both the correlated and uncorrelated contributions [see Fig.~\ref{FigScans}(b)], as well as a larger energy range [see Fig. \ref{FigScans}(c)]. 
This suggests that the additional `tripod' exchange paths introduced by the interlayer Cu atoms in Cu4 do not significantly change the excitations; instead they strengthen the interkagome correlations present in ZnCu3. 

These interkagome spin correlations are an important aspect of the Zn-claringbullite family and are likely due to the interkagome distance being shorter than the second nearest-neighbor one in the kagome planes. 
Exploring possible exchange models for Cu4 showed that a sizable antiferromagnetic interkagome exchange $J_c$ up to the order of 30\% of $J_1$, as well as a small further neighbor exchange $J_2$, are required to stabilize the magnetic structure. 
Although the sign of $J_c$ agrees with that previously suggested for barlowite  \cite{Jeschke2015}, we found that its magnitude relative to $J_1$ is greater.
In ZnCu3, the $Q$ positions of the magnetic responses could correspond to the [001] and [101] directions, and hence allude to interkagome spin correlations, shown to be the strongest by the zeroth-moment analysis. 
Despite the interkagome correlations in ZnCu3 favoring a three- rather than a two-dimensional system, a previous theoretical study using a coupled cluster method has shown that the kagome quantum spin liquid state can persist up to $J_c/J_1 \approx 15$\%, and it would be interesting to determine whether ZnCu3 fits into this QSL phase diagram \cite{Gotze2016}.

In comparison to ZnCu3, the powder and single crystal INS data of herbertsmithite also show two magnetic responses centered at $Q\approx0.67$~\Ang\ and $Q\approx1.3$~\Ang, which extend up to $\sim$2~meV \cite{Han2012} and $\sim$25~meV \cite{deVries2009}, respectively. 
Although part of the low-energy magnetic scattering was initially attributed to 15\% Cu impurity spins on the interlayer sites \cite{Han2016}, a variational Monte Carlo study showed it to be an intrinsic kagome response within the nearest-neighbor RVB model \cite{Zhang2020VariationalZnCu3OH6Cl2}. 
In our analysis of ZnCu3, both magnetic responses are attributed to the kagome layers through comparison with the parent material Cu4. 
However, the zeroth-moment analysis indicates that the correlations in ZnCu3 extend beyond the nearest-neighbor limit, suggesting a different flavor of QSL compared to herbertsmithite. 
This is further underlined by the difference in energy responses between the two systems: in herbertsmithite the excitations are underdamped \cite{Nilsen2013} while in ZnCu3 we find them to be overdamped.

The magnetic excitations of both Cu4 and ZnCu3 resemble those of the isostructural barlowite and its Zn-doped variant, Zn$_x$Cu$_{4-x}$(OD)$_6$FBr for $0\leq x <1$ \cite{Wei2017EvidenceAntiferromagnet}. For the $x\approx1$ sample, a finite sized gap of $\sim$0.65~meV is observed using $^{19}$F NMR \cite{Feng2017} and inelastic neutron scattering \cite{Wei2020}. In contrast, our INS data show ZnCu3 claringbullite to be gapless down to the lowest resolved energy transfer of $\sim$0.27~meV, but future NMR measurements are needed to verify this.

\section{Conclusions}
Our inelastic neutron scattering measurements show gapless ($\Delta<0.27$~meV), diffuse excitations arising from short-range magnetic correlations in ZnCu3 down to $T=1.7$~K suggesting the existence of a quantum spin liquid ground state. 
The analysis of the $Q$ and energy dependence of these correlations suggest a different ground state to that of herbertsmithite, widening the library of QSL types in experimental materials. 
ZnCu3 shows two inelastic magnetic responses that strongly resemble those of the pyrochlore-like Cu4 above its magnetic ordering temperature. 
In contrast, below $T_\mathrm{N}$, the two responses in Cu4 become gapped from each other and acquire a $\sim$0.5~meV zero-energy gap. 
Both responses are attributed to the kagome layers, with the low-energy one arising from interkagome correlations strengthened by the additional interlayer Cu in the parent material via the `tripod' interactions. 
Single crystal studies will be important in providing a more accurate exchange model for Cu4. 
Local probe measurements, such as NMR or muon spin relaxation studies, may also be helpful in determining whether the ZnCu3 state remains dynamic to mK temperatures. 
Furthermore, elemental analysis using anomalous x-ray diffraction could provide more accurate information on the extent of antisite disorder in ZnCu3.

\bigskip
\begin{acknowledgments}
This work was supported in part by the French Agence Nationale de la Recherche, Grant No.\ ANR-18-CE30-0022 LINK.  
The neutron diffraction and inelastic neutron scattering experiments were performed at the Institut Laue-Langevin (ILL) in Grenoble \cite{D2B_cbull, D20_cbull, Panther_cbull, IN5_cbull}, and experiments at the ISIS Neutron and Muon Source were supported by beamtime allocations from the UK Science and Technology Facilities Council (STFC \cite{LET_cbull}. We also acknowledge the ILL and ISIS deuteration facilities for provision of D$_2$O for the synthesis of deuterated samples. We thank Helen C. Walker for experimental support and Mike E. Zhitomirsky for discussions.. M.G. thanks the Department of Chemistry at UCL and the ILL for provision of a studentship.
\end{acknowledgments}

\end{document}


\title{Supplemental Material for \\ ``Magnetically ordered and kagome quantum spin liquid states in the Zn-doped claringbullite series''}

\author{M.~Georgopoulou} 
\affiliation{Institut Laue-Langevin, CS 20156, 38042 Grenoble Cedex 9, France}
\affiliation{Department of Chemistry, University College London, 20 Gordon Street, London, WC1H 0AJ, United Kingdom}

\author{B.~F\aa k}
\email{fak@ill.fr}
\affiliation{Institut Laue-Langevin, CS 20156, 38042 Grenoble Cedex 9, France}

\author{D. Boldrin} 
\affiliation{SUPA, School of Physics and Astronomy, University of Glasgow, Glasgow, G12 8QQ, United Kingdom}

\author{J. R. Stewart} 
\affiliation{ISIS Neutron and Muon Facility, Rutherford Appleton Laboratory, Science and Technology Facilities Council, Didcot OX11 0QX, UK}

\author{C. Ritter} 
\affiliation{Institut Laue-Langevin, CS 20156, 38042 Grenoble Cedex 9, France}

\author{E. Suard} 
\affiliation{Institut Laue-Langevin, CS 20156, 38042 Grenoble Cedex 9, France}

\author{J.~Ollivier} 
\affiliation{Institut Laue-Langevin, CS 20156, 38042 Grenoble Cedex 9, France}

\author{A.~S.~Wills} 
\affiliation{Department of Chemistry, University College London, 20 Gordon Street, London, WC1H 0AJ, United Kingdom}

\maketitle

\section{Sample synthesis}
Claringbullite, \ch{Cu4(OD)6FCl}, was synthesised by combining \ch{Cu2(OH)2CO3} (Sigma, 0.275~g, 1.25~mmol), \ch{CuCl2 * }2\ch{H2O} (Sigma, 0.215~g, 1.26~mmol), \ch{NH4F} (Alfa Aesar, 0.093~g, 2.50~mmol), \ch{HCl} (Sigma, 0.05~g, 37\% w.t. solution) and \ch{D2O} (10~mL) in a 15~mL Teflon-lined steel autoclave.

For Zn-claringbullite, with nominal stoichiometry \ch{ZnCu_3(OD)6FCl}, \ch{Cu2(OH)2CO3} (Aldrich, 0.250~g, 1.13~mmol), \ch{ZnCl2 * }$x$\ch{H2O} (Alfa, 0.190~g, 0.910~mmol), \ch{NH4F} (Alfa Aesar, 0.048~g, 1.29~mmol), \ch{HCl} (Sigma, 0.1~g, 37\% w.t.\ solution) and \ch{D2O} (10~mL) were placed in a 15~mL Teflon-lined steel autoclave. 

For both syntheses, the autoclaves were held at 200~$^\circ$C for 24~h and oven cooled to room temperature over 1~h. The products were washed {\textit{via}} centrifugation with \ch{D2O} (3 x 10~mL) and dried in an oven at 60~$^\circ$C for 4~h. Each synthesis produced approximately 0.2~g of material that were combined into a $\sim5$~g batch for Cu4 and $\sim7$~g batch for ZnCu3.

\section{Neutron diffraction}

\begin{figure}[!h]
\centering
\includegraphics[width=0.9\columnwidth]{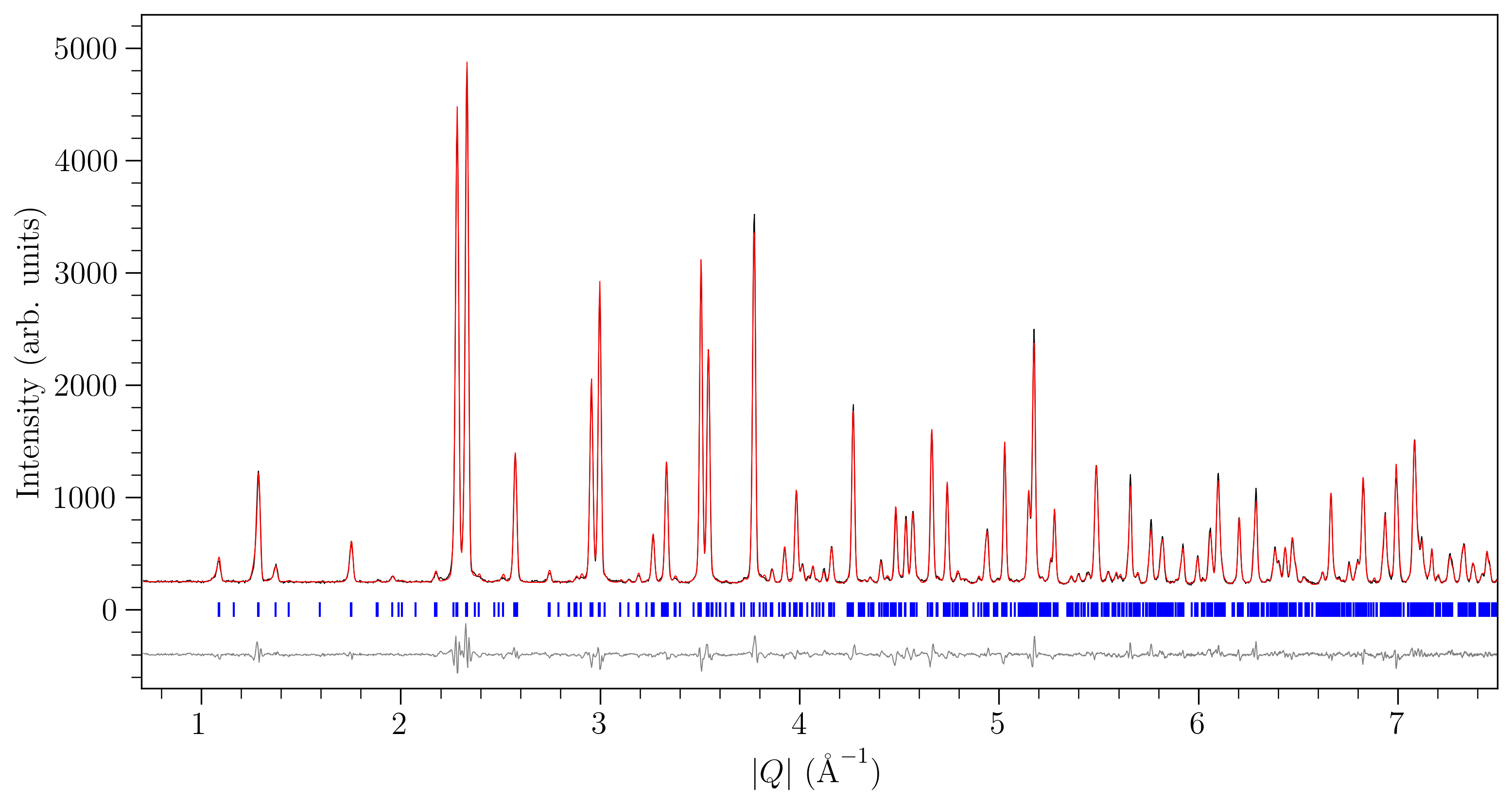}
\caption{Claringbullite, \clarid, D2B data collected at $T = 1.5$~K (black) with $\lambda = 1.595226$~\AA. Rietveld refinement (red) in the Pnma space group with 72 variables and goodness-of-fit parameters $\chi^2 = 2.69$ and $R_\mathrm{wp} = 4.17$. The difference between the experimental data and the fit is shown in grey and the peak positions are in blue.}
\label{Cu4RietveldRefinement}
\end{figure}

\begin{figure}[!h]
\centering
\includegraphics[width=0.9\columnwidth]{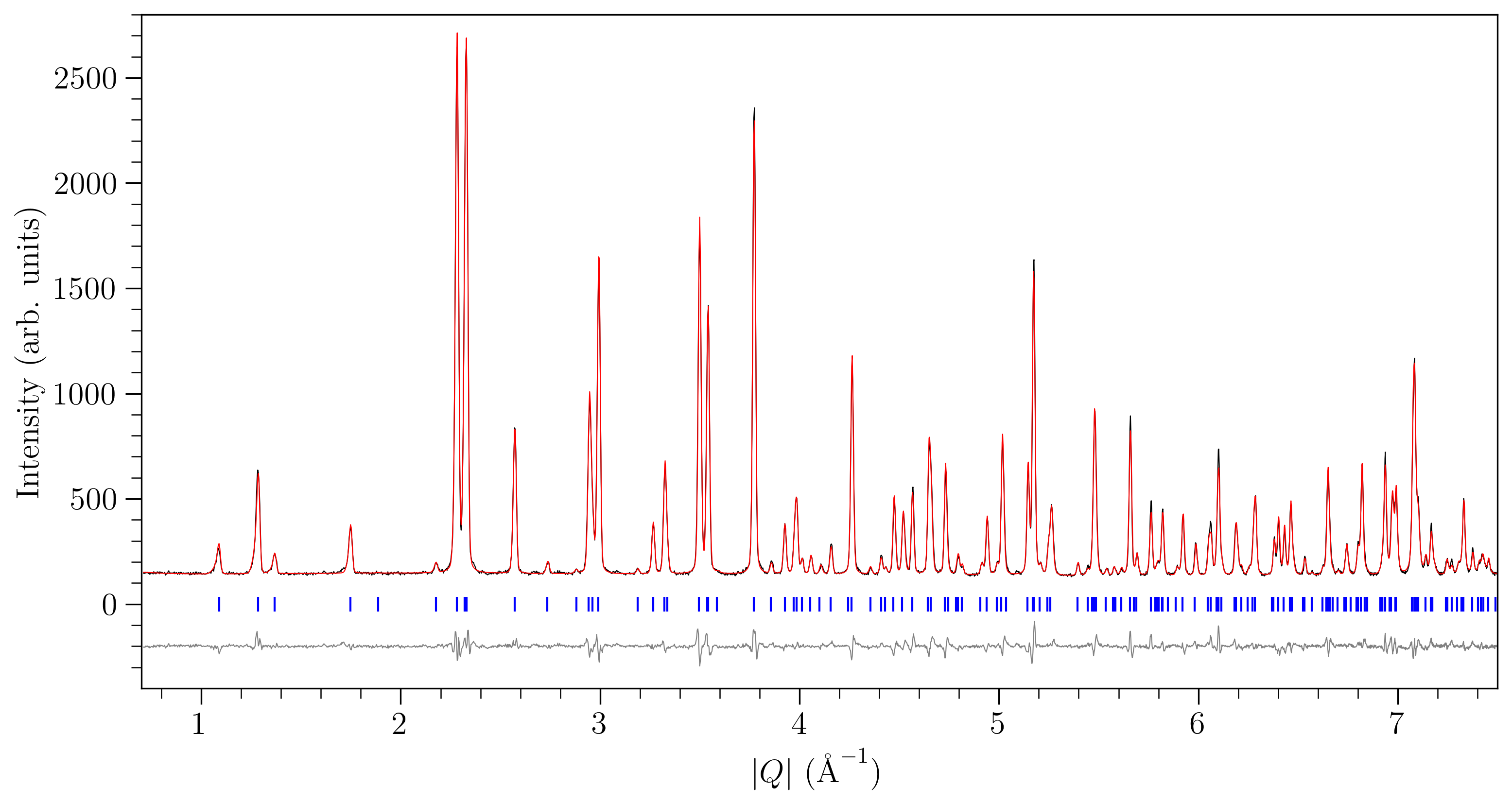}
\caption{Zn-claringbullite. Neutron diffraction data collected on D2B at $T = 1.5$~K (black) with $\lambda = 1.595226$~\AA. Rietveld refinement (red) in the $P6_3/mmc$ space group with 48 variables and goodness-of-fit parameters $\chi^2 = 2.28$ and $R_\mathrm{wp} = 4.60$. The D site refined to 0.96 showing good deuteration of the sample and was set to unity for the final refinement. The occupancy of the interlayer position refined to $\sim74$\% Zn occupation of the $2d$ site and $\sim8.5\%$ Cu occupation of the $6h$ site (see Table \ref{ZnCu3RietveldRefinementResults}). The difference between the experimental data and the fit is shown in grey and the peak positions are in blue. }
\label{ZnCu3RietveldRefinement}
\end{figure}

Neutron diffraction measurements were carried out on D2B for the deuterated claringbullite and Zn-doped claringbullite samples. The diffraction data measured at $T=1.5$~K and Rietveld refinements  are shown in Figs.~\ref{Cu4RietveldRefinement} and \ref{ZnCu3RietveldRefinement}, with the refined atomic parameters given in Tables~\ref{Cu4RietveldRefinementResults} and \ref{ZnCu3RietveldRefinementResults} for claringbullite and Zn-doped claringbullite, respectively. In the case of Zn-claringbullite, the Cu kagome site occupation refined close to unity and was subsequently fixed to 1. To estimate the site disorder at the interlayer site, a Cu atom was placed on the $6h$ site and its isotropic displacement parameter was fixed to that previously determined for Zn-doped barlowite at $T=1.5$~K using neutron scattering data \cite{Tustain2020}. The Zn $2d$ and Cu $6h$ site occupations were refined such that their sum was equal to the nominal stoichiometry of a divalent metal ion. The possibility of the interlayer Cu occupying the $12j$ site was explored, but there was no evidence from our refinements for Cu to occupy the $12j$ instead of the $6h$ site as previously proposed for Zn-doped barlowite \cite{Smaha2020}.

\begin{table}[hb!]
    \centering
    \begin{tabular}{p{2.5cm} p{2.5cm} p{2.5cm} p{2.5cm} p{2.5cm} p{2.5cm}} \hline \hline
    \multicolumn{6}{c}{Lattice parameters} \\ \hline
    $a$ (\AA) & $b$ (\AA) &  $c$ (\AA) & $\alpha$ (\degree) & $\beta$ (\degree) & $\gamma$ (\degree) \\ \hline
     11.5359(9) & 9.1510(7) &  6.6848(5) & 90 & 90 & 90 \\ \hline
     \multicolumn{6}{c}{Atomic parameters} \\ \hline 
     Atom  & Wyckoff site & $x$ & $y$ & $z$ & $U_\mathrm{iso}$ (\AA$^2$) \\ \hline
 Cu1 &  $4a$ &  0 & 0 & 0  & 0.00341(11) \\
 Cu2 & $8d$ & 0.2492(4) & 0.5121(3)  & 0.2470(5)  & 0.00238(57)  \\
 Cu3 & $4c$ & 0.1870(6) & 1/4 & 0.0592(6) & 0.00381(65)  \\
 F & $4c$ & 0.5007(11)  & 1/4 & 0.0041(13) & 0.00876(57)  \\
  Cl &  $4c$ &  0.3304(5) &  1/4 &  0.5049(6) & 0.00616(37)  \\
 O1 & $8d$ & 0.2961(5) & 0.0921(6)  & 0.00062(67)  & 0.0054(12)  \\
 O2   & $8d$ & 0.1022(5)  & 0.0919(5)  & 0.1989(7) &  0.00118(90) \\
 O3  & $8d$ & 0.4003(5)  & 0.5889(5)  & 0.3000(7)  & 0.00062(87)  \\
 D1  & $8d$ & 0.3765(7)  & 0.1342(7)  & 1.0030(8)  & 0.0150(11)  \\
 D2   & $8d$ & 0.0611(5)  & 0.1294(5)  & 0.3161(7) &  0.0139(11)  \\
 D3  & $8d$ & 0.4401(5)  & 0.6381(6) &  0.1912(7)  & 0.0173(11)  \\
    \hline \hline
   
\end{tabular}
    \caption{Claringbullite, \clarid. Lattice parameters, atomic positions and isotropic thermal parameters from the Rietveld refinement in the $Pnma$ space group using data collected on D2B ($\lambda=1.595226$~\AA) at $T=1.5$~K. All sites are fully occupied.}
    \label{Cu4RietveldRefinementResults}
\end{table}

\begin{table}[hb!]
    \centering
    \begin{tabular}{p{2.2cm} p{2.2cm} p{2.2cm} p{2.2cm} p{2.2cm} p{2.2cm} p{2.2cm} } \hline \hline
    \multicolumn{7}{c}{Lattice parameters} \\ \hline
    $a$ (\AA) & $b$ (\AA) &  $c$ (\AA) & $\alpha$ (\degree) & $\beta$ (\degree) & $\gamma$ (\degree) & \\ \hline
     6.65918(6) & 6.65918(6) &  9.17288(9) & 90 & 90 & 120 \\ \hline
    \multicolumn{7}{c}{Atomic parameters} \\ \hline
     Atom  & Wyckoff site & $x$ & $y$ & $z$ & Occupation & $U_\mathrm{iso}$ (\AA$^2$) \\ \hline
  Cu &  $6g$ &  0.5 & 0 & 0  & 1 & 0.0078 \\
 Zn & $2d$ & 1/3 & 2/3  & 3/4  & 0.74(2) & 0.0244  \\
  Cu & $6h$ & 0.3086(10) & 0.6172(19) & 3/4  & 0.085(8) & 0.0023 \cite{Tustain2020}  \\
 F & $2b$ & 0  & 0 & 3/4 & 1 & 0.0064  \\
 Cl &  $2c$ &  2/3 &  1/3 &  3/4 & 1 & 0.0075 \\
 O & $12k$ & 0.20184(10) & 0.79816(10) & 0.90838(11) & 1 & 0.0038  \\
 D  & $12k$ & 0.12414(9) & 0.87586(9) & 0.86584(11) & 1 & 0.0151 \\
    \hline 
    \multicolumn{7}{c}{Anisotropic displacement parameters (\AA$^2$)} \\ \hline 
  Atom &  U$_{11}$ & U$_{22}$ & U$_{33}$ & U$_{12}$ & U$_{13}$ & U$_{23}$ \\ \hline 
Cu & 0.0007(3) & 0.0007(3) & 0.0129(5) & 0.00034(15) & -0.0014(2) & -0.0027(4) \\
Zn & 0.016(3) & 0.016(3) & 0.0004(19) & 0.0081(13) & 0 & 0 \\
Cl & 0.0068(5) & 0.0068(5) & 0.0082(9) & 0.0034(2) & 0 & 0 \\
F & 0.0040(7) & 0.0040(7) & 0.0133(13) & 0.0020(3) & 0 & 0 \\
O & 0.0016(4) & 0.0016(4) & 0.0089(5) & 0.0009(5) &  -0.0010(2) & 0.0010(2)\\ 
D & 0.0132(5) & 0.0132(5) & 0.0217(6) & 0.0081(5) & -0.0007(2) & 0.0007(2) \\ \hline \hline
   
\end{tabular}
    \caption{Zn-claringbullite. Atomic positions and displacement parameters from Rietveld refinement in the $P6_3/mmc$ space group using D2B data ($\lambda=1.595226$~\AA) collected at $T=1.5$\,K. The D site refined to 0.96 showing good deuteration of the sample and was set to unity for the final refinement. The anisotropic displacements were refined for all atoms except the Cu on the $6h$ site. The refined stoichiometry is \znclariddis.}
    \label{ZnCu3RietveldRefinementResults}
\end{table}

\clearpage

\section{Bulk magnetometry}
\begin{figure}[!h]
\centering
\includegraphics[width=0.9\columnwidth, trim=4 4 4 4,clip]{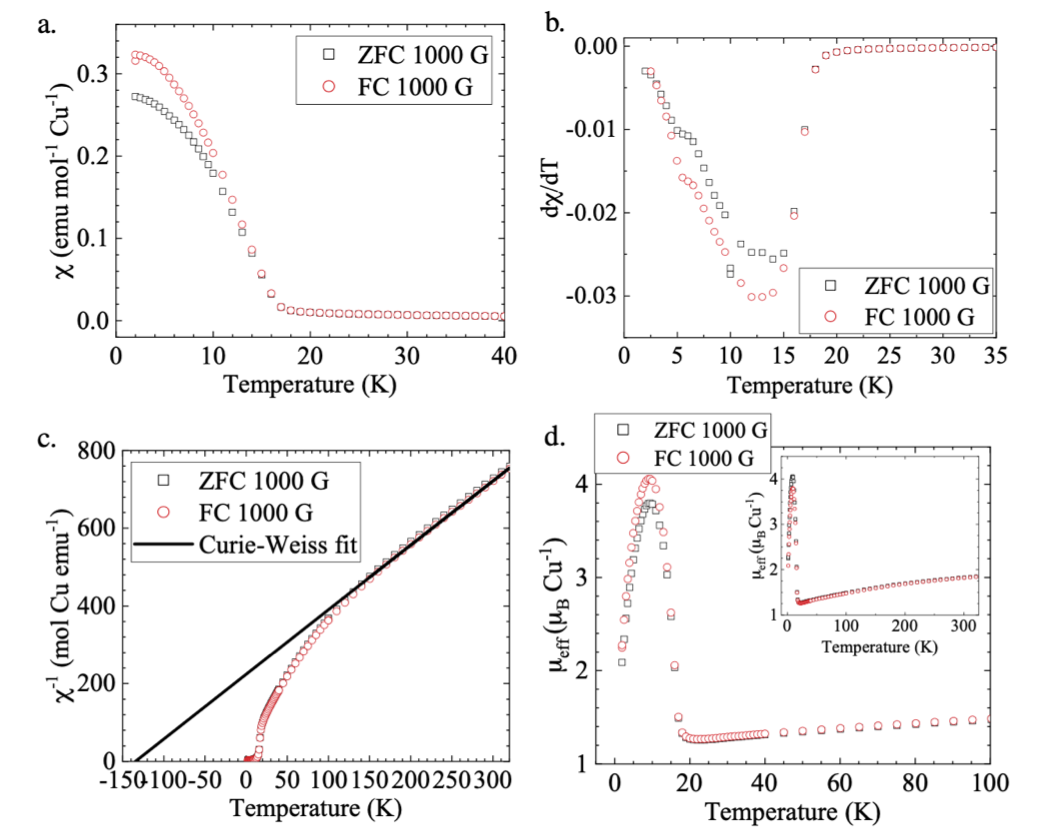}
   
\caption{Claringbullite. {\bf a.} Field-cooled (red) and zero-field-cooled (black) magnetic susceptibility data collected in a field of 1000~G. There is a transition at $T_\mathrm{N}$~=~17~K. {\bf b.} First derivative of the susceptibility to more clearly show the transition at 17~K. {\bf c.} Inverse susceptibility, $\chi^{-1}$, as a function of temperature, $T$, of field-cooled and zero-field-cooled data collected in 1000~G with a linear Curie-Weiss fit (black) between 150~K and 320~K that gives $\theta_\mathrm{W}=-136(3)$~K. {\bf d.} Effective magnetic moment, $\mu_{\mathrm{eff}}$, as a function of temperature, $T$, calculated using $\mu_{\mathrm{eff}}=\sqrt{8\chi T} $.}
\label{FigCu4Susceptibility}
\end{figure}

\begin{figure}[!h]
\centering
\includegraphics[width=0.9\columnwidth, trim=4 4 4 4,clip]{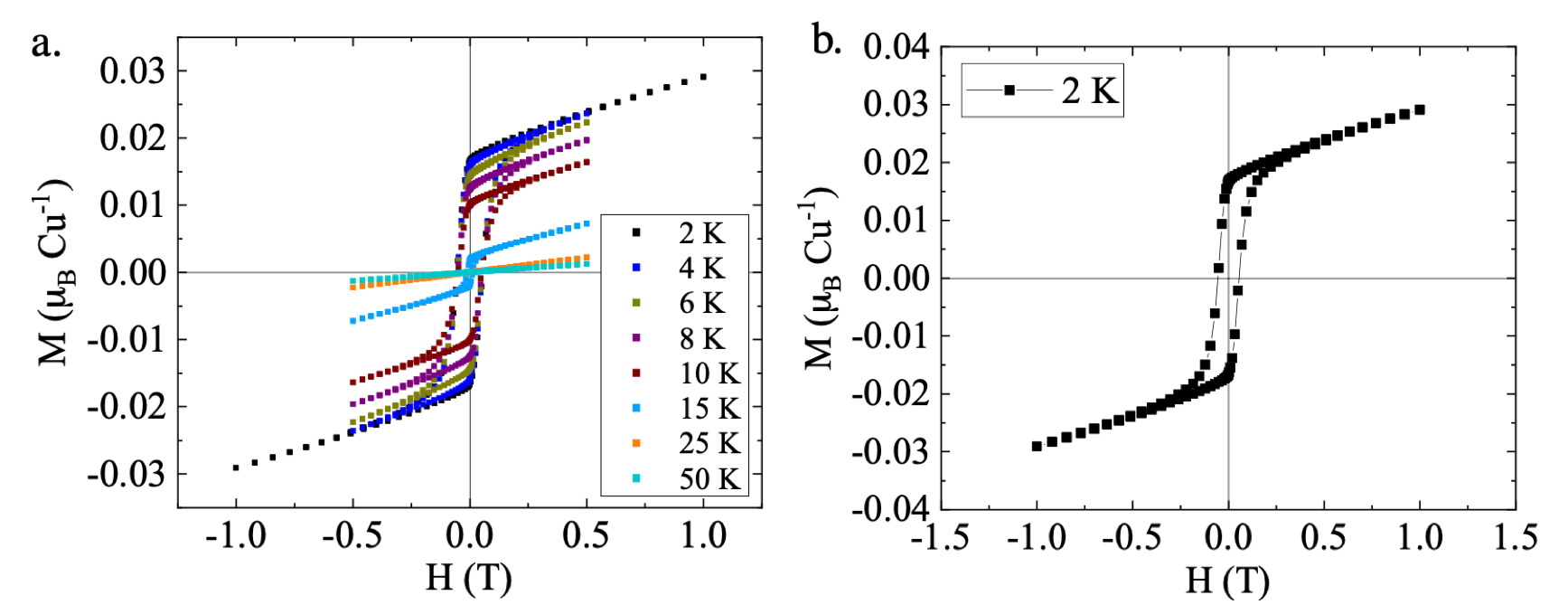}
   
\caption{Claringbullite. Magnetisation, $M$, as a function of field, $H$. \textbf{a.} Data between $T=2$~K and 50~K show a hysteresis loop opening at $T\leq 15$~K. \textbf{b.} At $T=2$~K the spontaneous moment is 0.017~\uB~Cu$^{-1}$ with a coercivity of 0.05~T.}
\label{FigCu4MvsH}
\end{figure}

\begin{figure}[!h]
\centering
\includegraphics[width=0.9\columnwidth]{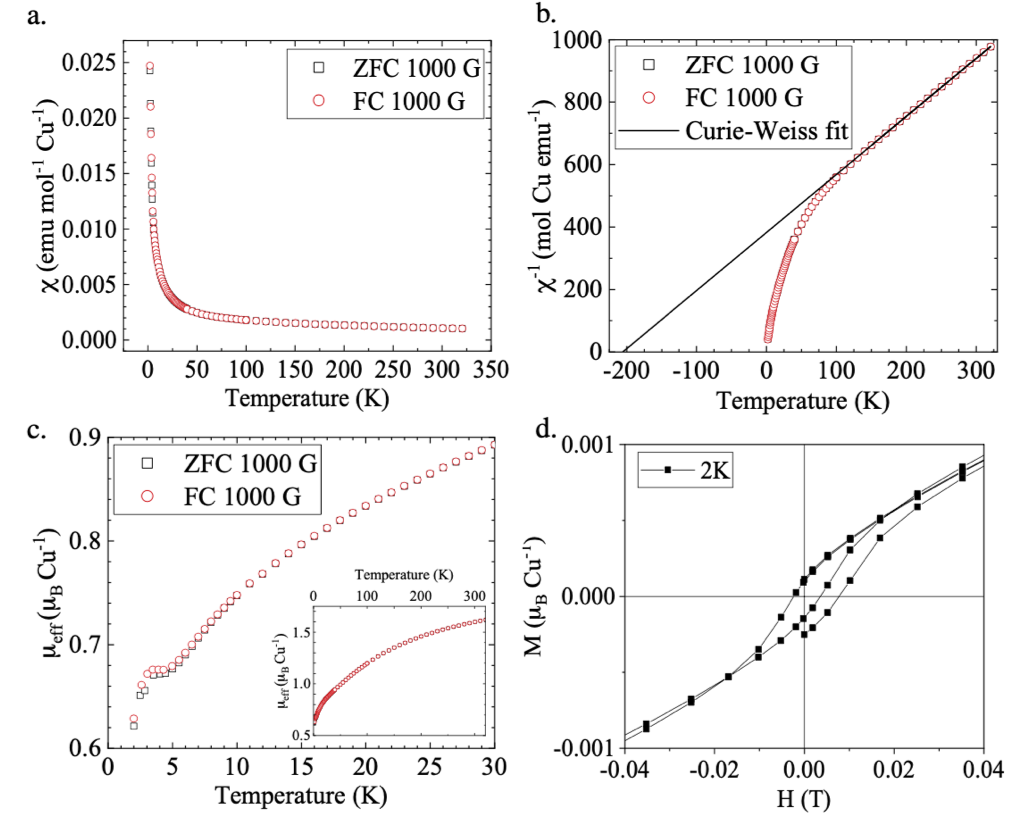}
   
\caption{Zn-claringbullite. \textbf{a.} Field-cooled (red) and zero-field-cooled (black) magnetic susceptibility data collected in a field of 1000 G. There is no transition down to 2\,K. \textbf{b.} Inverse susceptibility, ${\chi}^{-1}$, of field-cooled and zero-field-cooled data collected in 1000 G with a linear Curie-Weiss fit (black) between 150\,K and 320\,K that gives $\theta_\mathrm{W} = -206(1)$\,K. \textbf{c.} Effective magnetic moment, $\mu_{\textrm{eff}}$, as a function of temperature, $T$, calculated using $\mu_{\textrm{eff}}=\sqrt{8\chi T}$. \textbf{d.} Magnetisation, $M$, as a function of field, $H$, at $T=2$\,K showing a hysteresis loop with a spontaneous moment of $\sim2\times10^{-4}$~~\uB~Cu$^{-1}$ and a coercivity of $\sim3\times10^{-3}$\,T.}
\label{FigZnCu3Magnetometry}
\end{figure}

DC susceptibility measurements for claringbullite and Zn-claringbullite were carried out on a SQUID MPMS XL Quantum Design at the Institut N\'eel and a SQUID Quantum Design MPMS3 at the University of Glasgow. 
Magnetization measured as a function of temperature for claringbullite is shown in Fig.~\ref{FigCu4Susceptibility} and hysteresis loops at various temperatures are shown in Fig.~\ref{FigCu4MvsH}. Similarly for Zn-claringbullite, magnetization as a function of temperature and field are shown in Fig.~\ref{FigZnCu3Magnetometry}.

\clearpage

\section{Magnetic structure}
\begin{figure}[!hb]
\centering
\includegraphics[width=0.9\columnwidth]{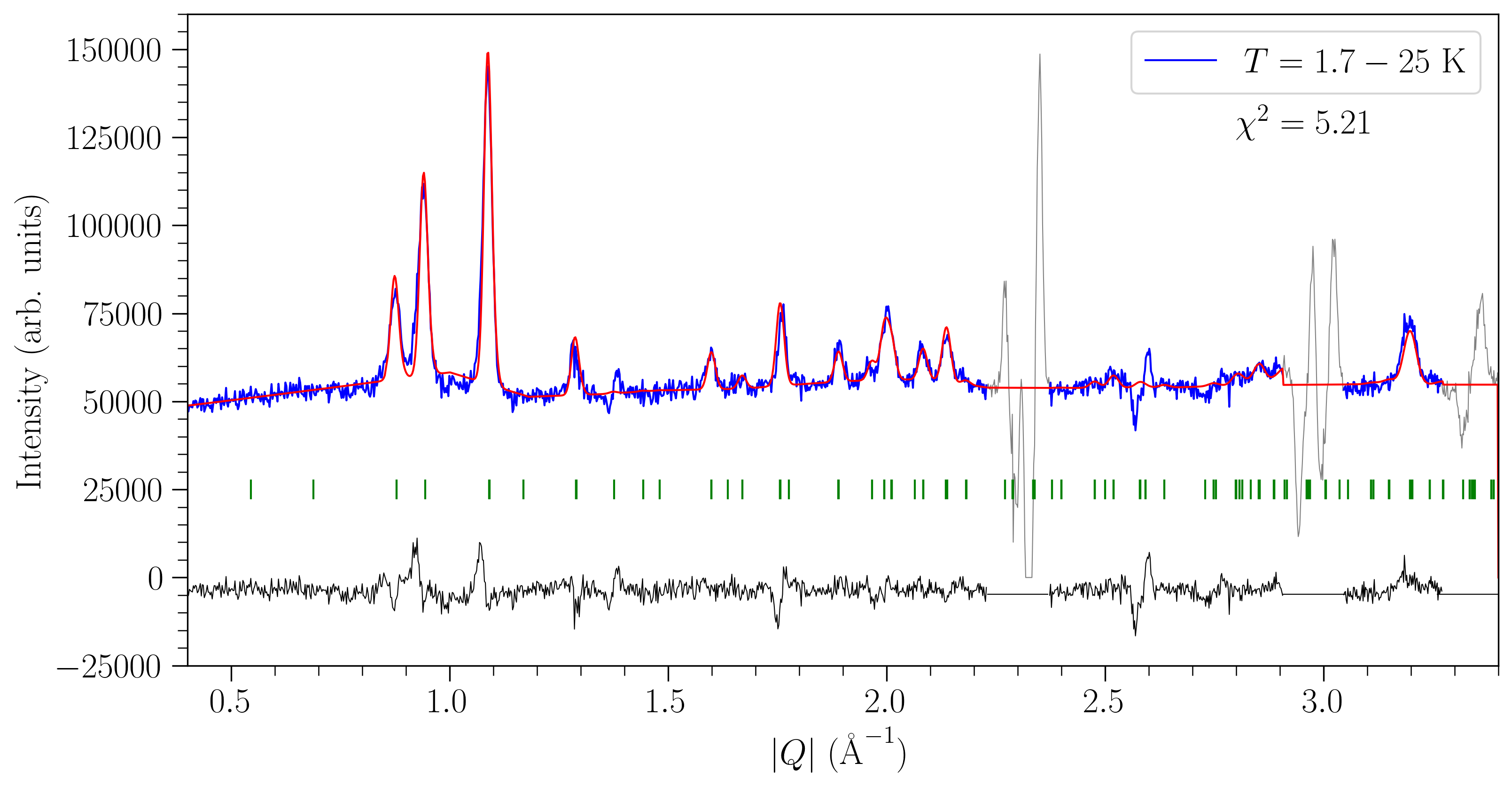}
   
\caption{Claringbullite. Magnetic structure refinement (red) with the IR $\Gamma_7$ using temperature subtracted data collected on D20 (blue). The magnetic Bragg peak positions are shown in green and the difference plot in black. The grey regions arise from the subtraction of nuclear peaks and were excluded from the refinements. The refined components of the magnetic moments are shown in Table \ref{cu4cbull_refined_moments}. }
\label{FigCu4MagStrRefinement}
\end{figure}

The magnetic structure of \clarid\ was refined using representation analysis in the space group $Pnma$ (no.~62) using the program SARAh \cite{Wills2000ASARAh} and the refinement is shown in Fig. \ref{FigCu4MagStrRefinement}. The basis vectors for $\Gamma_7$ are given in Tables \ref{cu4cbull_cu1_BVs}-\ref{cu4cbull_cu3_BVs}. The notation for the irreducible representations corresponds to that of Kovalev.

\begin{table*}[!h]
    \centering
    \begin{tabular}{c c c c c c} \hline \hline
     Atom number & Coordinates & {Basis vector} & {m$_a$} & m$_b$ &{m$_c$} \\ \hline
     Atom 1 & (0,~0,~0) & $\psi_1$ &  1 & 0 & 0 \\
    & & $\psi_2$ & 0 &  1 &  0  \\
     & & $\psi_3$ & 0 & 0 & 1  \\ 
      Atom 2 & (1/2,~1/2,~1/2) & $\psi_1$ & \minus 1 & 0 & 0 \\ 
      & & $\psi_2$ & 0 & 1 &  0  \\
     & & $\psi_3$ & 0 & 0 & 1  \\
     Atom 3 & (0,~1/2,~0) & $\psi_1$ &  1 & 0 & 0 \\
    & & $\psi_2$ & 0 & \minus 1 &  0  \\
     & & $\psi_3$ & 0 & 0 & 1  \\ 
     Atom 4 & (1/2,~0,~1/2) & $\psi_1$ & \minus 1 & 0 & 0 \\
    & & $\psi_2$ & 0 & \minus 1 &  0  \\
     & & $\psi_3$ & 0 & 0 & 1  \\ 
     
     \hline \hline
   
\end{tabular}
    \caption{The basis vectors of the $\Gamma_7$ irreducible representation of the space group $Pnma$ with $\mathbf{k}=(0,~0,~0)$ for the Cu1 $4a$ site. m$_a$, m$_b$ and m$_c$ represent the components with respect to the crystallographic axes.}
    \label{cu4cbull_cu1_BVs}
\end{table*}

\begin{table*}[!h]
    \centering
    \begin{tabular}{c c c c c c} \hline \hline
      Atom number & Coordinates & {Basis vector} & {m$_a$} & m$_b$ &{m$_c$} \\ \hline
       Atom 1 & (0.247,~0.496,~0.249) & $\psi_1$ & 1 & 0 & 0 \\ 
      & & $\psi_2$ & 0 & 1 &  0  \\
     & & $\psi_3$ & 0 & 0 & 1  \\
     Atom 2 & (0.747,~0.004,~0.251) & $\psi_1$ & \minus 1 & 0 & 0 \\
    & & $\psi_2$ & 0 & 1 &  0  \\
     & & $\psi_3$ & 0 & 0 & 1  \\ 
     Atom 3 & (0.753,~0.996,~0.751) & $\psi_1$ & 1 & 0 & 0 \\
    & & $\psi_2$ & 0 & \minus 1 &  0  \\
     & & $\psi_3$ & 0 & 0 & 1  \\ 
     Atom 4 & (0.253,~0.504,~0.743) & $\psi_1$ & \minus 1 & 0 & 0 \\
    & & $\psi_2$ & 0 & \minus 1 &  0  \\
     & & $\psi_3$ & 0 & 0 & 1 \\ 
     Atom 5 & (0.753,~0.504,~0.751) & $\psi_1$ &  1 & 0 & 0 \\
    & & $\psi_2$ & 0 & 1 &  0  \\
     & & $\psi_3$ & 0 & 0 & 1  \\
     Atom 6 & (0.253,~0.996,~0.749) & $\psi_1$ & \minus 1 & 0 & 0 \\
    & & $\psi_2$ & 0 & 1 &  0  \\
     & & $\psi_3$ & 0 & 0 & 1  \\
     Atom 7 & (0.247,~0.004,~0.249) & $\psi_1$ & 1 & 0 & 0 \\
    & & $\psi_2$ & 0 & \minus 1 &  0  \\
     & & $\psi_3$ & 0 & 0 & 1  \\
      Atom 8 & (0.747,~0.496,~0.251) & $\psi_1$ & \minus 1 & 0 & 0 \\
    & & $\psi_2$ & 0 & \minus 1 &  0  \\
     & & $\psi_3$ & 0 & 0 & 1  \\
       \hline \hline
   
\end{tabular}
    \caption{The basis vectors of the $\Gamma_7$ irreducible representation of the space group $Pnma$ with $\mathbf{k}=(0,~0,~0)$ for the Cu2 $8d$ site. m$_a$, m$_b$ and m$_c$ represent the components with respect to the crystallographic axes.}
    \label{cu4cbull_cu2_BVs}
\end{table*}

\begin{table*}[!htb]
    \centering
    \begin{tabular}{c c c c c c} \hline \hline
    Atom number & Coordinates & {Basis vector} & {m$_a$} & m$_b$ &{m$_c$} \\ \hline
       Atom 1 & (0.191,~1/4,~0.051) & $\psi_1$ & 1 & 0 & 0 \\ 
      & & $\psi_2$ & 0 & 0 &  1  \\
     Atom 2 & (0.691,~1/4,~0.449) & $\psi_1$ & \minus 1 & 0 & 0 \\
    & & $\psi_2$ & 0 & 0 &  1  \\
     Atom 3 & (0.809,~3/4,~0.949) & $\psi_1$ & 1 & 0 & 0 \\
    & & $\psi_2$ & 0 &0 & 1  \\
     Atom 4 & (0.309,~3/4,~0.551) & $\psi_1$ & \minus 1 & 0 & 0 \\
    & & $\psi_2$ & 0 & 0 &  1  \\
      \hline \hline
   
\end{tabular}
    \caption{The basis vectors of the $\Gamma_7$ irreducible representation of the space group $Pnma$ with $\mathbf{k}=(0,~0,~0)$ for the Cu3 $4c$ site. m$_a$, m$_b$ and m$_c$ represent the components with respect to the crystallographic axes.}
    \label{cu4cbull_cu3_BVs}
\end{table*}

\begin{table*}[!htb]
    \centering
    \begin{tabular}{c c c c c c} \hline \hline
     Atom & Wyckoff site & Coordinates & {m$_a$} ($\mu_\mathrm{B}$) & m$_b$ ($\mu_\mathrm{B}$) &{m$_c$} ($\mu_\mathrm{B}$) \\\hline
     Cu1 & $4a$ &  (0,~0,~0) &   0.266 & 0 & 0\\
             & & (1/2,~1/2,~1/2)  & \minus0.266 & 0 & 0\\
             & & (0,~1/2,~0) & 0.266 & 0 & 0\\
             & & (1/2,~0,~1/2) & \minus0.266 & 0 & 0\\
    \hline
    Cu2 & $8d$ &(0.247,~0.496,~0.249) & 0.353 & 0.120 & 0 \\
        &  & (0.747,~0.004,~0.251) & \minus0.353 & 0.120 & 0 \\
        & & (0.753,~0.996,~0.751)  & 0.353 & \minus0.120 & 0 \\
        & & (0.253,~0.504,~0.743) & \minus0.353 & \minus0.120 & 0 \\
       & &  (0.753,~0.504,~0.751) & 0.353 & 0.120 & 0 \\
       & &  (0.253,~0.996,~0.749) & \minus0.353 & 0.120 & 0 \\
       & &  (0.247,~0.004,~0.249) & 0.353 & \minus0.120 & 0 \\
       & & (0.747,~0.496,~0.251)  & \minus0.353 & \minus0.120 & 0 \\

    \hline
    Cu3 &  $4c$ & (0.191,~1/4,~0.051) & 0.491 & - &  0.153 \\
           &  & (0.691,~1/4,~0.449) & \minus0.491 & - &  0.153 \\
           & &  (0.809,~3/4,~0.949) & 0.491 & - &  0.153 \\
          &  & (0.309,~3/4,~0.551) & \minus0.491 & - &  0.153 \\
       \hline \hline
   
\end{tabular}
    \caption{The refined components of the magnetic moments for the equivalent positions of the three Cu sites (Cu1, Cu2 and Cu3) along the $a$, $b$ and $c$ crystallographic directions of the $Pnma$ unit cell. }
    \label{cu4cbull_refined_moments}
\end{table*}

\clearpage
\section{Spin wave model of claringbullite compatible with inelastic neutron scattering data}
\begin{figure}[!hb]
\centering
\includegraphics[width=0.6\columnwidth, trim=3 1 10 7,clip]{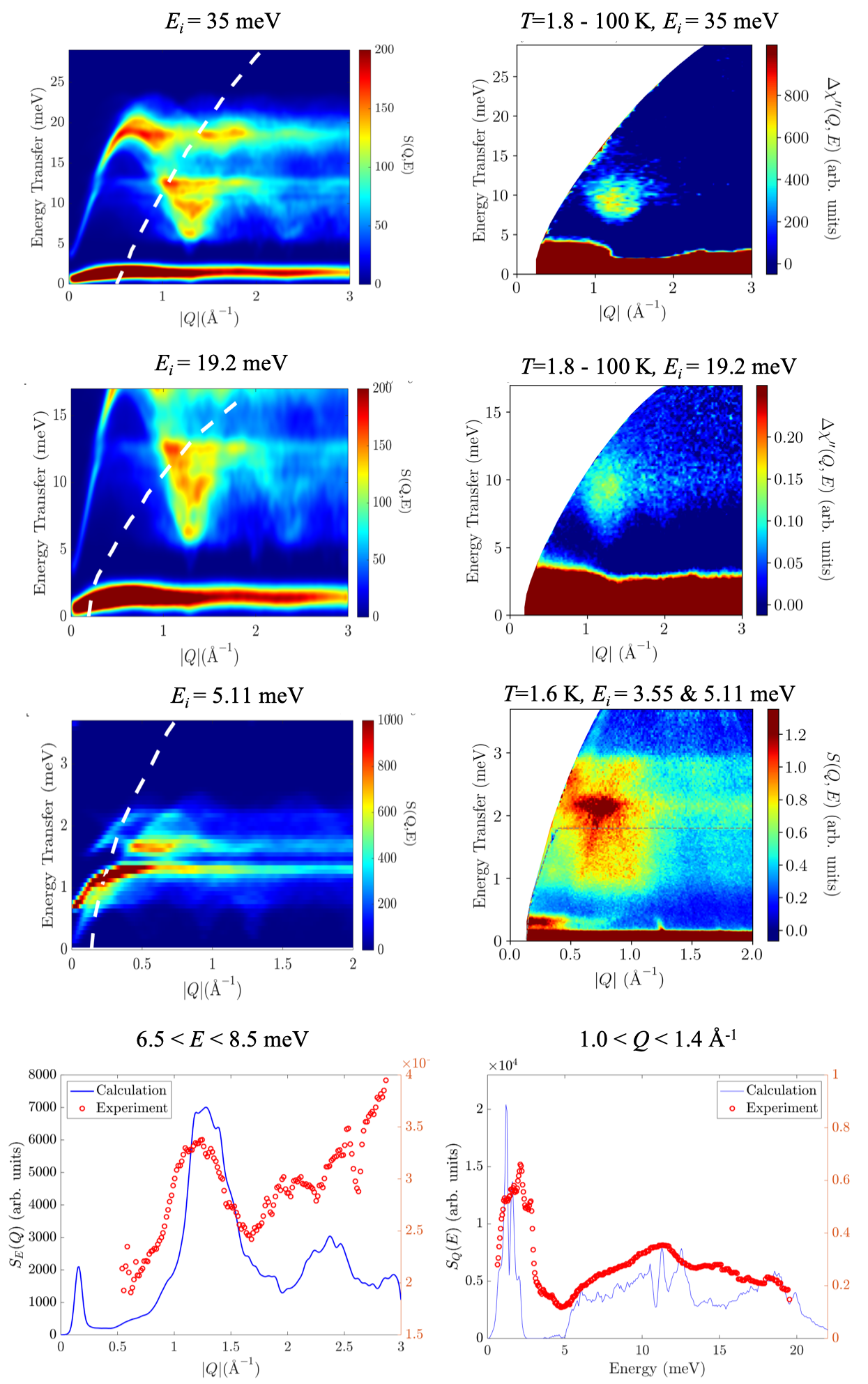}
   
\caption{Claringbullite calculated spin wave spectra ({\bf left.}) using the exchange interactions in Table~I in the main text, compared to the experimental data ({\bf right.}). The directions of the Dzyaloshinskii-Moriya interactions are as described in the main text. The dashed white lines are the kinematic windows for the corresponding incident neutron energies. $S_E(Q)$ and $S_Q(E)$ scans of the calculated (blue) compared to the experimental (red) data are shown at the {\bf bottom.}}
\label{FigSpinWModels}
\end{figure}

Powder averaged spin wave excitations were calculated in SpinW \cite{Toth2015LinearStructures} for \clarid\ using the exchange interactions given in Table~I in the main text and the magnetic structure shown in Fig.~2 in the main text. These are compared to the experimental inelastic neutron scattering data collected on PANTHER and IN5 at the ILL in Fig.~\ref{FigSpinWModels}.

